 \documentstyle[epsfig,twocolumn]{iso98}		

\newcommand{\thepapername}{}
\newcommand{\psection}[1]{\section*{#1}
\addtocounter{section}{1}
\setcounter{subsection}{0}
\setcounter{subsubsection}{0}
}

\makeindex

\def\cm-2{$cm^{-2}$}

\def	\gtsim	{\lower.5ex\hbox{$\; \buildrel > \over \sim \;$}} 
\def	\ltsim	{\lower.5ex\hbox{$\; \buildrel < \over \sim \;$}} 

\def	\beq	{\begin{displaymath}}	
\def	\eeq	{\end{displaymath}}	

\def	\cm	{{\rm \,cm}}

\def\ltsim{$\buildrel < \over \sim $}
\def\gtsim{$\buildrel > \over \sim $}

\newcommand   {\ga}    {\mbox{\rlap{\hbox{\lower4pt\hbox{$\sim$}}}\hbox{$>$}}}
\newcommand   {\la}    {\mbox{\rlap{\hbox{\lower4pt\hbox{$\sim$}}}\hbox{$<$}}}

\def\ltsima{$\; \buildrel <\over \sim \;$}
\def\simlt{\lower.5ex\hbox{\ltsima}}
\def\gtsima{$\;\buildrel > \over \sim \;$}
\def\simgt{\lower.5ex\hbox{\gtsima}}

\def\h13cop{\hbox{H$^{13}$CO$^{+}$}}

\DeclareRobustCommand{\ion}[2]{%
\relax\ifmmode
\ifx\testbx\f@series
{\mathbf{#1\,\mathsc{#2}}}\else
{\mathrm{#1\,\mathsc{#2}}}\fi
\else\textup{#1\,{\mdseries\textsc{#2}}}%
\fi}

\newcounter{ioncnt}
\newcommand{\markcite}[1]{\relax}
\newcommand{\placefigure}[1]{}

\newcommand{\tablewidth}[1]{}

\begin{document}

\setlength{\parindent}{0pt}
\setlength{\parskip}{ 10pt plus 1pt minus 1pt}
\setlength{\hoffset}{-1.5truecm}
\setlength{\textwidth}{ 17.1truecm }
\setlength{\columnsep}{1truecm }
\setlength{\columnseprule}{0pt}
\setlength{\headheight}{12pt}
\setlength{\headsep}{20pt}

\pagestyle{myheadings}		
\pagenumbering{roman}           

\setcounter{tocdepth}{0}


\twocolumn[
	\begin{center}
	\vspace{2cm}
	\section*{\bf FIRBACK FAR INFRARED SURVEY WITH ISO: \\	
	       	      DATA REDUCTION, ANALYSIS AND FIRST RESULTS}

\setcounter{section}{0}
\setcounter{figure}{0}
\setcounter{table}{0}
\setcounter{footnote}{0}
\setcounter{equation}{0}
	\vspace{0.5cm}	

{\bf H.~Dole\index{Dole, H.|textbf}$^1$, G.~Lagache\index{Lagache, G.}$^1$, 
J-L.~Puget\index{Puget, J.-L.}$^1$, R. Gispert\index{Gispert, R.}$^1$, 
H. Aussel\index{Aussel, H.}$^2,^7$, 
F.R. Bouchet\index{Bouchet, F.R.}$^3$,  \\
\bf C. Ciliegi\index{Ciliegi, C.}$^4$, 
D.L. Clements\index{Clements, D.L.}$^5$, C.J. Cesarsky\index{Cesarsky, C.J.}$^2$, 
F.-X. D\'esert\index{Desert@D\'esert, F.X.}$^6$,  D. Elbaz\index{Elbaz, D.}$^2$,  
 A. Franceschini\index{Franceschini, A.}$^7$,  \\ 
\bf B. Guiderdoni\index{Guiderdoni, B.}$^3$, M. Harwit\index{Harwit, M.}$^8$, 
R. Laureijs\index{Laureijs, R.}$^9$,  D. Lemke\index{Lemke, D.}$^{10}$,  
R. McMahon\index{McMahon, R.}$^{11}$, A.F.M. Moorwood\index{Moorwood, A.F.M.}$^{12}$, \\ 
\bf S. Oliver\index{Oliver, S.}$^{13}$, W.T. Reach\index{Reach, W.T.}$^{14}$, 
M. Rowan-Robinson\index{Rowan-Robinson, M.}$^{13}$, M. Stickel\index{Stickel, M.}$^{10}$}\\ \vspace{2mm} 
$^1$Institut d'Astrophysique Spatiale, Orsay, France\\
$^2$Service d'Astrophysique, CEA/DSM/DAPNIA Saclay, France \\
$^3$Institut d'Astrophysique de Paris, France\\
$^4$Osservatorio Astronomico di Bologna, Italy\\
$^5$Cardiff University, UK\\
$^6$Laboratoire d'Astrophysique, Observatoire de Grenoble, France\\
$^7$Osservatorio Astronomico di Padova, Italy\\
$^8$511 H.Street S.W., Washington, DC 20024-2725\\
$^9$ISOC ESA, VILSPA, Madrid, Spain\\
$^{10}$MPIA, Heidelberg, Germany\\
$^{11}$Institute for Astronomy, University of Cambridge, UK\\
$^{12}$ESO, Garching, Germany\\
$^{13}$Imperial College, London, UK\\
$^{14}$IPAC, Pasadena, CA, USA

\vspace{1cm}
\end{center}
]
	
\addcontentsline{toc}{section}{FIRBACK survey with ISO: data reduction, 
analysis and first results \\ 
H. Dole et al.}	
\thispagestyle{mfkheadings}

\begin{abstract}
FIRBACK  is one of the deepest cosmological surveys performed in the
far infrared, using ISOPHOT. We describe this survey, its data reduction 
and analysis. We present the maps of fields at 175\,$\mu m$.
We point out some first results: source identifications with radio and 
mid infrared, and source counts at 175\,$\mu m$. These two results 
suggest that half of the FIRBACK  sources are probably at redshifts greater 
than 1. We also present briefly the large follow-up program.
  \vspace {5pt} 

  Key~words: ISO; ISOPHOT; far infrared survey; cosmology; galaxy evolution

\end{abstract}

\psection{1. INTRODUCTION}
The discovery of the Cosmic Far Infrared Background Radiation (CFIBR) (\cite{puget96}, 
\cite{fixsen98}, \cite{hauser98}, \cite{lagache99a}) using COBE data and constraints 
on its spectrum, demonstrates that about two thirds of the light emitted by galaxies 
integrated over all redshifts has been processed by dust and released at FIR and 
submm wavelengths. This background is in line with models using strong evolution of galaxies.\\
Studying the population of galaxies radiating mostly at long wavelengths as a function 
of redshift is thus one of the important but difficult observational tasks of today 
in the field of galaxy formation and evolution (see Puget \& Lagache, this volume).\\
In this context, FIRBACK, which is a cosmological survey dedicated to a study 
of the Cosmic Far Infrared Background and galaxies contributing to it, 
is one important step.\\

 \begin{figure*}[!ht]
  \begin{center}
   \leavevmode
\vspace{1cm}
  \centerline{\epsfig{file=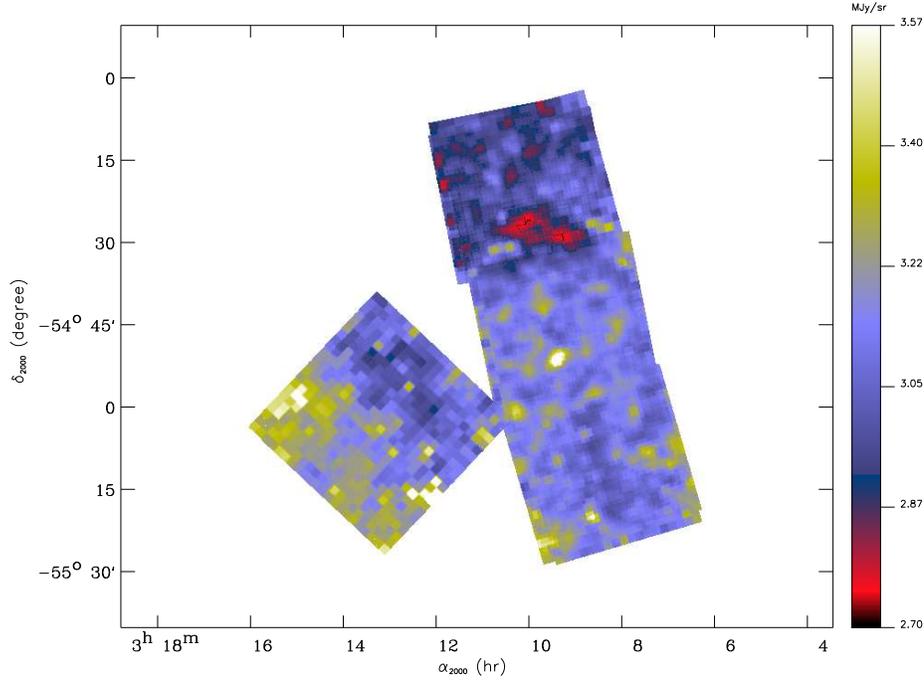,width=10.0cm}}
  \vspace{-5cm}
  \end{center}
  \caption{\em Map of FIRBACK  Marano fields}
  \label{fig:map_marano}
\end{figure*}

 \begin{figure*}[!ht]
  \begin{center}
   \leavevmode
\vspace{1cm}
  \centerline{\epsfig{file=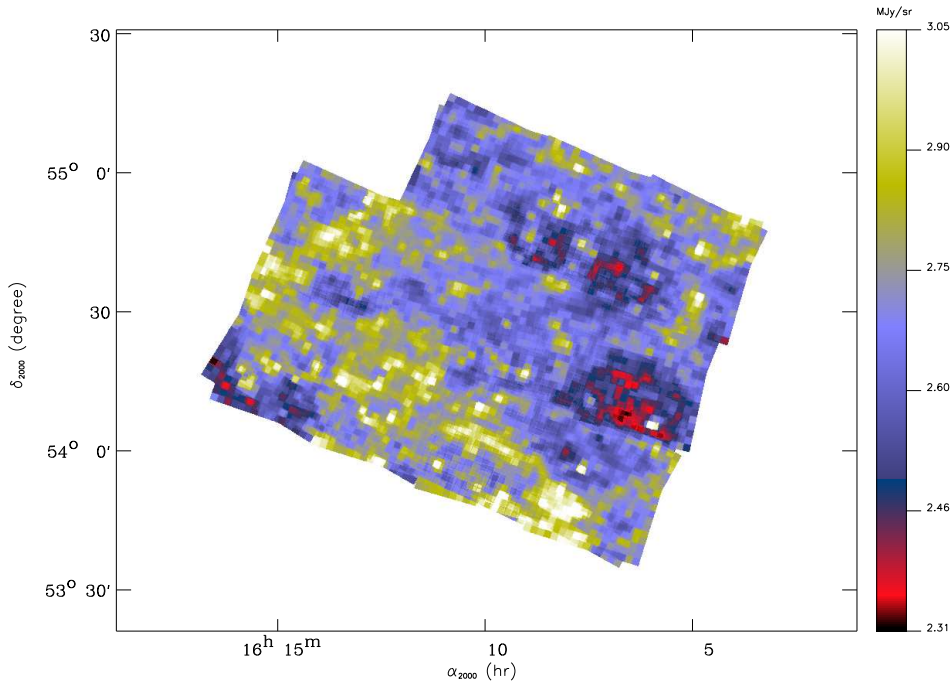,width=10.0cm}}
  \vspace{-5cm}
  \end{center}
  \caption{\em Map of FIRBACK  ELAIS N1 field}
  \label{fig:map_elaisn1}
\end{figure*}

\psection{2. DATA SET}
\label{sec:data_set}

FIRBACK  is a survey of 3 main fields (described in Table~\ref{tab:fields}) covering 
4 square degrees using ISOPHOT.
All fields have been observed in raster mode (AOT P22) with the \verb+C200+ 
detector and \verb+C_160+ filter centered at $\lambda = 175\, \mu m$. We 
have also additional ISO observations with ISOPHOT and ISOCAM (see below).

\begin{table}[htb]
  \caption{\em Fields of the FIRBACK  survey.}
 \label{tab:fields}
  \begin{center}
    \leavevmode
    \footnotesize
    \begin{tabular}[h]{lccccc}
      \hline \\[-5pt]
      Field Name & $\alpha_{2000}$	&  $\delta_{2000}$	&  \it{l}	& \it{b}	& area   \\[+5pt]
	         & h min  	 	& deg min		& deg		& deg		& sq deg \\
      \hline \\[-5pt]
      Marano   & 03 11 & -54 45 & 270 & -52 & 1\\
      ELAIS N1 & 16 11 & +54 25 &  84 & +45 & 2\\
      ELAIS N2 & 16 36 & +41 05 &  65 & +42 & 1\\
      \hline \\
      \end{tabular}
  \end{center}
\end{table}

\vspace{6mm}
\begin{center} {\it FIRBACK  Marano} \end{center}
\vspace{-5mm}
This area (map Figure~1) is composed of four individual 
fields, called Marano 1, 2, 3 and 4, which have been observed four times. 
The redundancy goes from 32 (in overlaping regions) to 1 (on the edges) with 
an average of 16, and the oversampling is optimal in both Y and Z directions 
in Marano 2, 3 and 4 (half pixel offset between rasters). Marano 1 covers 
$31.5' \times 31.5'$, and Marano 2, 3 and 4 about $77' \times 26'$. 
Integration time per sky pixel is 256 s on average. 
\textit{Additional observations:} 
in Marano 1 we have one P25 absolute measurement, five small rasters at 90\,$\mu m$ 
(not yet reduced); in the complete Marano field, we have a 15\,$\mu m$ ISOCAM survey.

 \begin{figure}[!hb]
  \begin{center}
   \leavevmode
\vspace{1cm}
  \centerline{\epsfig{file=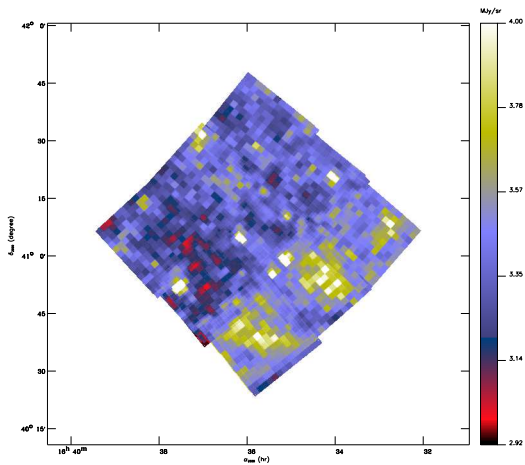,width=7.0cm}}
  \end{center}
  \vspace{-2cm}
  \caption{\em Map of FIRBACK  ELAIS N2 field}
  \label{fig:map_elaisn2}
\end{figure}

\begin{center} {\it FIRBACK  ELAIS N1} \end{center}
\vspace{-5mm} 
This area (map Figure~2) is composed of eleven individual fields, 
observed 2 times with an offset of less than a pixel, and the redundancy goes from 
16 (in overlaping regions) to 1 (on the edges) with an average of 8. ELAIS N1 covers 
about 2 square degrees. Integration time per sky pixel is 128 s on average. Note that 
the northern FIRBACK  fields have been choosen to cover some corresponding ELAIS fields 
at other wavelengths (see Rowan-Robinson et al., this volume).

\begin{center} {\it FIRBACK  ELAIS N2} \end{center}
\vspace{-5mm} 
This area (map Figure~3) is composed of nine individual fields, 
observed 2 times with an offset of one pixel, so the redundancy is 8 on average, 
without oversampling. ELAIS N2 covers about $57' \times 57'$.

\psection{3. DATA REDUCTION}
\label{sec:data_reduction}

\subsection{From Raw Data to AAP}
\label{sec:data_raw_analysis}
We used PIA\footnote{PIA is a joint development by the ESA Astrophysics Division 
and the ISOPHOT Consortium} version 7.2.2 (Phot Interactive Analysis, 
\cite{gabriel97} ) to reduce data from ERD (Edited Raw Data, the lowest level) 
to AAP (Auto Analysis Product Data, level where flux density and positions 
are available): \\
- ERD to SRD (Signal per Ramp Data): linearisation, deglitching, fit ramps to first order\\
- SRD to SCP (Signal per Chopper Plateau Data): deglitching, no drift correction \\
- SCP to AAP: no reset interval correction (reset times are identical), subtraction of 
dark current, power calibration using interpolation between 2 FCS.\\
We then used our own IDL routines to correct for transients induced by strong glitches, 
and to perform a flat correction using observational redundancies.
\subsection{Map Projection}
\label{sec:map_projection}

Our basic data projection is performed in two steps: (1) geometrical grid computation and 
(2) projection of
the signal.\\
- The first step is the most time consuming. We read all AAP
data of one complete field, and create a large grid in $(\alpha, \delta)$ with tunable 
pixel sizes (usually 10 arcsec); this
step has to be performed just one time.\\
- The second step just acts to restore the grid, to coadd individual raster data onto 
the grid by distributing the pixel signal in the area of one pixel, and by updating a 
weighting map.\\
This method is well-suited to large amounts of data, in terms of sky surface or 
number of co-added rasters with different roll angles. 
Figures~\ref{fig:map_marano},~\ref{fig:map_elaisn1} and~\ref{fig:map_elaisn2} show such maps.

\psection{4. RESULTS}

\subsection{Source Detection}
Source detection is directly performed on final co-added maps, using eye recognition to
discriminate point sources from cirrus structure.
We then computed the flux by aperture photometry with radii choosen to minimize 
noise in source detection $R_{min} = 60''$ and $R_{max} = 92''$ (\cite{puget99}).

\subsection{Catalogs and Identifications}
\vspace{-3mm}
Sources in our catalog have a S/N ratio greater than 3, and photometric accuracy is about 30 \%. 
Fluxes are in the range 75 mJy to 1.6 Jy, with an average of 252 mJy and a median of 175 mJy.\\
In order to constrain both Spectral Energy Distribution (SED) and position of these sources, 
we looked for identifications: 
Table~\ref{tab:sources} gives a summary of detected sources and identifications with 
ISO or ground-based radio telescopes.\\
In the {\it Marano Field}, we performed a smaller 15\,$\mu m$ survey using ISOCAM 
(\cite{desert}), which covers about 50 \% of the 175\,$\mu m$ survey surface.
Radio identifications are performed using Australia Telescope Compact Array 
available data (\cite{grup97}) or preliminary data of our program being reduced, 
that is why 8 identifications is a lower limit.\\
The situation is different in northern fields {\it ELAIS N1 \& N2} because 
the ELAIS consortium performed observations using ISO at $\lambda =$ 15\,$\mu m$, 
$\lambda =$ 90\,$\mu m$ (Rowan-Robinson et al., this volume), using the VLA at $\lambda =$ 20\,$ cm$ 
(\cite{ciliegi98}), and at other wavelengths.
Most of the 31 sources with 90\,$\mu m$ counterpart have a color ratio 
$\frac{F_{175\,\mu m}}{F_{90 \,\mu m\, PHOT}} \ge 2$, which has to be 
compared to local sources ($z \le 0.1$) where 
$\frac{F_{175\,\mu m}}{F_{100 \,\mu m\, IRAS}} \simeq 1$ (\cite{stickel98}), 
suggesting that a significant proportion of FIRBACK  sources are large-redshift 
sources (see discussion below).
Most of NED\footnote{The NASA/IPAC Extragalactic Database (NED) 
is operated by the Jet Propulsion Laboratory, California Institute 
of Technology, under contract with the National Aeronautics and 
Space Administration} identifications are FIRST (\cite{becker95}) 
or IRAS sources.

\begin{table}
  \caption{\em Number of FIRBACK  sources and identifications}
 \label{tab:sources}
  \begin{center}
    \leavevmode
    \footnotesize
    \begin{tabular}{lcccc}
      \hline \\[-5pt]
      Field Name	&175\,$\mu m$ 		& 15\,$\mu$m		&90\,$\mu$m		&20\,cm			\\[+5pt]
      \hline \\[-7pt]
    $F_{\nu}(mJy) \ge $ & 75			& 1			& 60			& 0.2 \\
      \hline \\[-5pt]
      Marano   & 78  & 29 & no data  & $\ge 8$  \\
      ELAIS N1 & 113 & 19 & 18 & 20  \\
      ELAIS N2 & 85  & no data & 13 & 22  \\
      \hline \\
      \end{tabular}
\vspace{-1cm}
  \end{center}
\end{table}

\subsection{Source Counts}
\begin{figure}[!ht]
  \begin{center}
    \leavevmode
\vspace{6mm}
  \centerline{\epsfig{file=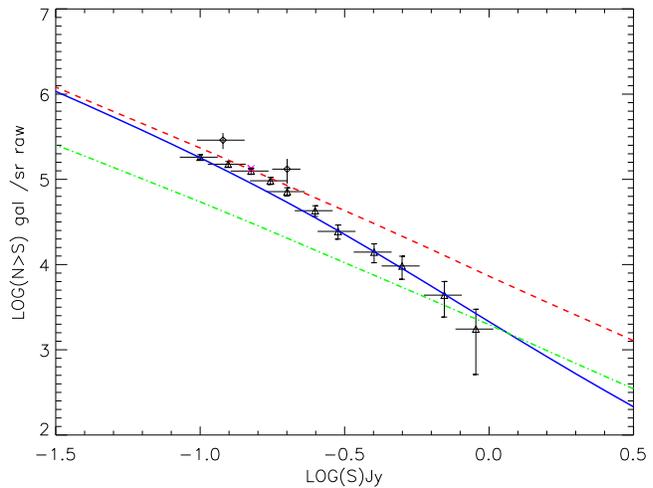,width=7.0cm}}
 \vspace{-3.5cm}
  \end{center}
  \caption{\em Source counts with FIRBACK  data in Marano 1, 2, 3, 4, ELAIS N1 and ELAIS N2. 
  These counts are not corrected for incompleteness. Data: diamonds: 
  Marano 1 (Puget et al. 1999); triangles: ELAIS N1, N2 and Marano 2 3 4; 
  Lockman Hole data (Kawara et al. 1998): purple star at 150 mJy, superimposed 
  on our counts. Models: Guiderdoni et al. 1998: blue line: model E; Franceschini et al. 1998: 
  red dash-dash: with evolution; green dash-dot: without evolution}
  \label{fig:counts}
\end{figure}

We present raw source counts in Figure~\ref{fig:counts}, using sources coming 
from FIRBACK  ELAIS N1, N2 and Marano 2 3 4 fields (208 sources brighter 
than 100 mJy in less than 3 square degrees). Raw counts means here that 
\textit{counts are not corrected for incompleteness}.
For $log(S \ge 300 \, mJy)$, counts are not expected to change a lot 
due to incompleteness correction, so the steep slope of 2.20 is a minimum. 
We also plotted Marano 1 raw counts (\cite{puget99}) which are higher 
by a factor 2 but compatible within uncertainties, probably because based 
on 22 sources in a smaller area (0.25 square degree), Kawara et al. (1998) 
data in the Lockman Hole (difficult to see in the plot because their count 
at 150 mJy gives the same number as ours), models with and without 
evolution from Franceschini et al. (1998), and model E from Guiderdoni et al. (1998).
Compared to extrapolated IRAS counts, FIRBACK  counts are about 
20 times higher at 200 mJy. Our counts are not compatible with 
non-evolution models and need strong evolution. Model with 
evolution from Franceschini et al. (1998) predicts smoother slopes than 
observed, whereas model E from Guiderdoni et al. (1998) fits roughly 
the slope even if lower than observed.\\

\psection{5. FOLLOW-UP \footnote{extended FIRBACK collaboration for follow-up 
also includes: D. Benford, P. Cox, M. Dennefeld, G. Helou, R. Mac Mahon, 
A. Omont, F. Pajot, T. Phillips, D. Scott, F. Viallefond}}
\vspace{-6mm}
An intensive multiwavelength follow-up program is being performed. \\
In the {\it Marano field}, we are following-up at $\lambda$ = 20 cm 
using the Australia Telescope Compact Array in compact configurations 
(data being reduced) and extended configuration (program to submit)(Dole). 
In the optical range, one source has been observed with the NTT (Dennefeld).\\
In the {\it ELAIS N1 \& N2 fields} at millimetre wavelengths,
 we are following-up using the IRAM 30 m antenna and Plateau de 
 Bure Interferometer (Omont \& Guiderdoni). Submillimetre observations 
 in N1 are scheduled at JCMT (Scott) and in April at CSO (Lagache), 
 and optical follow-up of sources is performed at Palomar (Reach).

\psection{6. DISCUSSION ON COSMOLOGICAL IMPLICATIONS}
\subsection{Population}
The steep slope of the number counts ($\ge$ 2.2) cannot be accounted 
for by the effect of the K-correction (ratio, at a given wavelength, 
of the emitted flux and the observed flux with a spectral redshift) 
if no cosmological evolution is present as can be seen in Figure~\ref{fig:counts}.\\
The Guiderdoni et al. (1998) model which gives rather a good fit to our 
counts presents a strong evolution for the star burst component 
($\propto (1+z)^5$) and an ULIRG (Ultra Luminous Infrared Galaxies) 
component. In this model, the far infrared background comoving 
luminosity per unit volume increases by a factor of 20 from $z = 0$ to $z = 2.5$.\\
This model also predicts a redshift distribution for sources 
brighter than 100 mJy at 175\,$\mu m$. The redshift distribution 
peaks at $z = 0.9$ with a median of $z = 1.0$, about half of 
the sources lie at redshift between 1 and 2, and about 10\% above $z = 2$. 
Furthermore, this model is compatible with the far infrared 
backgound detected in the COBE data. Sources detected in our 
survey account for 3\% of this background at 175\,$\mu m$, 
and the model predicts that 90\% of the background is made 
of sources brighter than 2 mJy; this predicted value is probably 
too low, because our slope is higher than the model, and our counts 
are not yet corrected for incompleteness.
\vspace{-5mm}
\subsection{Identifications}
\begin{figure}
  \begin{center}
    \leavevmode
  \centerline{\epsfig{file=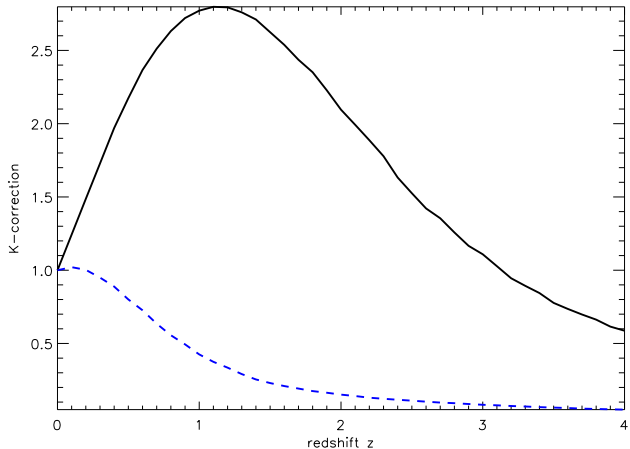,width=9.0cm}}
 \vspace{-1.5cm}
  \end{center}
  \caption{\em K-correction at 175\,$\mu m$ (black line) and 90\,$\mu m$ (blue dash-dash line) as a function of redshift for a starburst galaxie whose SED peaks at 75\,$\mu m$}
  \label{fig:kcorrection}
\end{figure}

\vspace{-5mm}
If these galaxies are star burst galaxies with a SED peaking at 75\,$\mu m$, 
a typical source observed close to our detection limit 
($F_{175 \,\mu m} = 100$\,mJy) has a luminosity 
$1.2 \times 10^{12} L_\odot$ (in a $h=0.5$ and $\Omega_0=1$ cosmology). \\
The same source will have $F_{15\,\mu m} \simeq 1$\,mJy and thus is 
easily detectable with ISOCAM. One of the differences between 
mid- and far-infrared observations is that the K-correction 
allows observation of galaxies with redshifts in general below 
1.4 (because of the PAH feature's cutoff) in the mid infrared; 
no such strong limit exists in the far infrared.
We have only 25\% detections at 15\,$\mu m$, implying that more 
than half of the sources are at $z > 1.4$ 
(redshift at which 15\,$\mu m$ flux drops sharply). \\
The expected 90\,$\mu m$ flux for this source is $F_{90\,\mu m} \simeq 40$\,mJy, 
and thus not easily detectable in the ELAIS survey. Futhermore, 
the K-correction at 90 and 175\,$\mu m$ as function 
of redshift (Figure~\ref{fig:kcorrection}) shows that between 
redshifts $z=1$ and $z=2$ the ratio 
$\frac{K-corr_{175\,\mu m}}{K-corr_{90\,\mu m}}$ grows from 
6 to 14: at $z=2$ a similar galaxy has 14 times more flux 
at $175\,\mu m$ than at $90\,\mu m$. \\
The non-thermal emission is known to be strongly correlated 
with the FIR emission for starburst galaxies (e.g \cite{helou85}), 
even if Sanders \& Mirabel (1996) show an important scatter in the correlation. 
With ELAIS VLA identifications (42 sources), we compute 
$q = \log(F_{\nu \,FIR}) - \log(F_{\nu \, 1.4 GHz})$ 
(we take $F_{175\,\mu m}$ for $F_{\nu \,FIR}$ and not 
a combination of $F_{60\,\mu m}$ and $F_{100\,\mu m}$) 
the FIR to 1.4 GHz log ratio. For 70~\% of our sample 
$q < 2.3$, and $2.3 \le q < 3$ for 30~\%; the dispersion 
is significant and must be taken into account for deep radio observations.\\

 \vspace{-0.5cm}

\psection{7. CONCLUSION}
We presented the FIRBACK  175\,$\mu m$ survey, data reduction and analysis. 
Identifications of sources at various wavelengths, color ratios and 
source counts suggest that half of the sources are at redshifts 
greater than 1. Following-up these sources in order to have 
better positional accuracies, and SED constraints is now the 
key for understanding this population of dust enshrouded galaxies.\\
Additional information, images and papers are available online 
at \textbf{\verb+http://wwwfirback.ias.fr+}

\psection{ACKNOWLEDGMENTS}
We would like to thank M.A. Miville-Desch\^enes for his courtesy 
in permitting use of his image printing program, and C. Gabriel for 
fruitful discussions at ESA's ISO Data Centre (IDC) at Vilspa, Spain.

\onecolumn

\cleardoublepage




\begin{thebibliography}{}

\bibitem[\protect\astroncite{Becker et al.}{1995}]{becker95}
Becker, R.H., et al.  1995, ApJ, 450, 559

\bibitem[\protect\astroncite{Ciliegi et al.}{1998}]{ciliegi98}
Ciliegi, C. et al.  1998, astro-ph/9805353

\bibitem[\protect\astroncite{Condon,}{1992}]{condon92}
Condon, J.J., 1992, Annu. Rev. A. A. 30, 575

\bibitem[\protect\astroncite{D\'esert et al.}{1998}]{desert}
D\'esert, F-X. et al.  1998 A\&A, astro-ph/9809004

\bibitem[\protect\astroncite{Elbaz et al.}{1998}]{elbaz98}
Elbaz, D. et al.  1998, astro-ph/9807209 

\bibitem[\protect\astroncite{Fixsen et al.}{1998}]{fixsen98}
Fixsen, D.J., et al.   1998, ApJ, 508, 123 

\bibitem[\protect\astroncite{Franceschini et al.}{1998}]{franc98}
Franceschini, A., et al. 1998, MNRAS 296, 709 

\bibitem[\protect\astroncite{Gabriel et al.}{1997}]{gabriel97}
Gabriel, C. et al.  1997, Proc. of the ADASS VI conf., ASP Conf. Ser., 
Vol.125, p108

\bibitem[\protect\astroncite{Gruppioni et al.}{1997}]{grup97}
Gruppioni, C. et al.  1997, MNRAS, 286, 470

\bibitem[\protect\astroncite{Guiderdoni et al.}{1998}]{guider98}
Guiderdoni, B. et al., 1998, MNRAS 295, 877

\bibitem[\protect\astroncite{Hauser et al.}{1998}]{hauser98}
Hauser, M.G., et al.  1998 ApJ, 508, 25

\bibitem[\protect\astroncite{Helou et al.}{1985}]{helou85}
Helou, G., et al.  1985, ApJ, 298, L7

\bibitem[\protect\astroncite{Kawara et al.}{1998}]{kawara98}
Kawara, K., et al.  1998, A\&A 336, L9

\bibitem[\protect\astroncite{Lagache et al.}{1999}]{lagache99a}
Lagache, G., et al.  1999, A\&A, in press, astro-ph/9901059

\bibitem[\protect\astroncite{Puget et al.}{1996}]{puget96}
Puget, J.L., et al.  1996, A\&A 308, L5

\bibitem[\protect\astroncite{Puget et al.}{1999}]{puget99}
Puget, J.L., et al.  1999, A\&A, in press, astro-ph/9812039



\bibitem[\protect\astroncite{Sanders \& Mirabel}{1996}]{sanders96}
Sanders, D.B., Mirabel, I.F., 1996, Annu. Rev. Astron. Astrophys. 34, 749

\bibitem[\protect\astroncite{Stickel et al.}{1998}]{stickel98}
Stickel M., et al.  1998, A\&A, 336, 116

\end{thebibliography}
\end{document}